\documentclass[11pt,letterpaper]{article}
\usepackage[T1]{fontenc}
\usepackage[utf8]{inputenc}
\usepackage{lmodern}
\usepackage[margin=1in]{geometry}
\usepackage{amsmath,amssymb,amsthm,bm}
\usepackage{graphicx,booktabs,array,float}
\usepackage{natbib}
\usepackage{microtype}
\usepackage{setspace}
\usepackage[font=small,labelfont=bf]{caption}
\usepackage{xcolor}
\usepackage{hyperref}
\hypersetup{colorlinks=true,linkcolor=blue!45!black,citecolor=blue!45!black,urlcolor=blue!45!black,
 pdftitle={New Demand Economics: Education, Demand Upgrading, and Structural Change},
 pdfauthor={Fenghua Wen; Xieyu Yin; Chufu Wen}}
\newtheorem{proposition}{Proposition}

\newtheorem{corollary}{Corollary}
\theoremstyle{definition}
\newtheorem{definition}{Definition}
\newcommand{\logit}{\operatorname{logit}}
\newcommand{\Lam}{\Lambda}
\newcommand{\dd}{\mathrm{d}}

\newcommand{\figurenote}[1]{\par\vspace{3pt}\begin{minipage}{\linewidth}\footnotesize\setstretch{1.05}\textit{Notes:} #1\end{minipage}}
\begin{document}
\hypersetup{pageanchor=false}
\begin{titlepage}
\centering
\vspace*{0.3in}
{\LARGE\bfseries New Demand Economics: Education, Demand Upgrading, and Structural Change\par}
\vspace{0.25in}
{\large Fenghua Wen$^{1,2,*}$\qquad Xieyu Yin$^1$\qquad Chufu Wen$^1$\par}
\vspace{0.12in}
{\small $^1$School of Business, Central South University\par
$^2$Business School, Hunan Institute of Technology\par}
\vspace{0.2in}
{September 2026\par}
\vspace{0.27in}
\begin{minipage}{0.93\textwidth}
\begin{center}\textbf{Abstract}\end{center}
\setstretch{1.12}
Education can change what households buy as well as what workers produce. We study this demand channel in a two-sector growth model. Education shifts spending toward an education-sensitive bundle, drawing labor into the sector that supplies it. Learning lowers its relative price. When substitution between the bundles is strong enough, the price decline raises its expenditure share and reinforces the initial shift. The balance between this feedback and technological catch-up determines the long-run allocation. Weak feedback gives a unique stable configuration. Strong feedback can sustain two stable configurations with different expenditure shares, so the long-run outcome depends on initial productivity. A sufficiently large, attainable increase in education removes the low-share configuration. This transition can occur through demand alone, without a direct productivity effect of education. Early consumption commitments can also change the economy's destination, even though their budget share eventually vanishes. The growth and welfare consequences depend on learning gained in the expanding sector and learning forgone in the other. We derive the net learning benefit that a planner would attach to reallocation.
\vspace{0.18in}
\end{minipage}
\vfill
\begin{minipage}{0.95\textwidth}
\footnotesize\setstretch{1.03}
$^*$Corresponding author: \href{mailto:wfh@amss.ac.cn}{wfh@amss.ac.cn}. Other authors: \href{mailto:yinxieyu@csu.edu.cn}{yinxieyu@csu.edu.cn}; \href{mailto:wenchufu@sina.com}{wenchufu@sina.com}.\par\smallskip
An earlier English version circulated as \textit{New Demand Economics} (arXiv:2510.17121, October 2025). Our Chinese-language article, \textit{Demand Hierarchy, Education, and High-Quality Growth}, published in \textit{Systems Engineering---Theory \& Practice}, presents an earlier and simpler formulation of the core theoretical argument. The present paper substantially reworks and extends the theoretical framework. The authors acknowledge support from the National Natural Science Foundation of China (Grant No.\ 72131011).
\end{minipage}
\end{titlepage}
\hypersetup{pageanchor=true}
\setcounter{footnote}{0}
\section{Introduction}
Learning to read changes what a household can obtain from a book. Education can therefore change spending even when income and prices stay the same. If that spending draws resources into book production, it may also affect what producers learn and the future price of books. This paper asks when such a change in consumption can produce a lasting change in the structure of production.

Economic history offers a useful starting point. In de Vries's account of the \emph{industrious revolution}, households increased their demand for purchased goods while reallocating work toward market production \citep{DeVries1994}. Consumption was part of the process through which markets expanded. The history of books adds the role of acquired skills. \citet{Buringh2009} document the long expansion of manuscript and printed-book production in Western Europe. They describe changes in readership and discuss falling prices and higher literacy as likely sources of growth after the mid-fifteenth century. These accounts bring household capabilities, purchases, and production into the same historical question.

We study that question in a two-sector growth model. Households consume a basic bundle and an \emph{education-sensitive bundle}, whose value to them rises with education. We call an increase in the expenditure share of the latter \emph{demand upgrading}. The distinction can apply to different uses or qualities within an industry, as well as to goods from different industries. Labor moves between the two sectors to meet household demand. Learning depends on sectoral production, and the sector that lags behind in productivity benefits from technological catch-up.

To see the mechanism, start with an economy in which all spending is discretionary. Education shifts expenditure toward the education-sensitive bundle at given income and prices. The resulting shift of labor raises learning in that sector relative to the other sector and lowers its relative price. When the elasticity of substitution exceeds one, households respond by spending a larger share on the now cheaper bundle. Demand and learning then reinforce one another. Catch-up limits the persistence of productivity gaps.

In this economy, a balanced-growth configuration has constant expenditure shares and a constant productivity ratio. The main result shows when different such configurations can persist. If feedback is weak relative to catch-up, every initial productivity ratio converges to the same balanced-growth configuration. If feedback is strong enough, two stable configurations can coexist: one with a low expenditure share on the education-sensitive bundle and one with a high share. An unstable configuration separates them. Relative productivity below the unstable threshold leads to the low-share configuration; relative productivity above it leads to the high-share configuration. A sufficiently large increase in education removes the low-share configuration, provided that the required education level is attainable. The economy then moves toward the high-share configuration. This transition can arise through demand alone, even when education has no direct effect on productivity.

Consumption commitments can affect this process long after they cease to absorb much of the household budget. Early spending to meet these commitments directs labor, and hence learning, toward the basic sector. When two stable configurations coexist, this early allocation can determine the economy's long-run destination. The model thus shows how economies with the same current education level can retain different production patterns because of their earlier consumption constraints.

The growth consequences depend on learning opportunities in both sectors. A shift toward the education-sensitive bundle adds learning in one sector and gives up learning in the other. We derive the resulting growth comparison and the value a planner places on this reallocation. These results distinguish the demand channel of education from its direct productive effect and identify the learning benefits that enter a welfare assessment.

\subsection{Relation to the literature}
The model connects three lines of research. The first studies how income and relative prices reshape demand as economies grow. \citet{Kongsamut2001} characterize structural transformation alongside balanced aggregate growth, and \citet{Herrendorf2014} review the main mechanisms. \citet{Boppart2014} and \citet{Comin2021} distinguish income and substitution effects in structural change. We add education as a source of demand shifts at a given income. Combining this shift with sectoral learning yields an explicit condition for multiple balanced-growth configurations.

The second examines how demand can sustain growth. In \citet{Matsuyama2002}, productivity growth makes goods affordable to more households, expanding markets and supporting further productivity growth. \citet{Foellmi2006} connect demand and income distribution to innovation, while \citet{Aoki2002} study demand saturation and the creation of new demand. Our focus is on the capabilities households bring to consumption. Education changes what they value, and the interaction of demand, learning, and catch-up determines whether that change produces gradual adjustment or a transition between stable configurations.

The third concerns education and consumption. \citet{Michael1973} studies education's role in nonmarket production. \citet{Stigler1977} explain how accumulated capital and productive capacities can change consumption under stable underlying tastes. \citet{Buera2012} link skills to the expansion of services, and \citet{Alan2018} show that a classroom intervention can affect intertemporal choice. In our model, an education-dependent preference weight summarizes the effect of capabilities, experience, and valuation on demand. The size of this response is an empirical question for the particular bundles under study.

On the supply side, we draw on learning by doing \citep{Arrow1962} and the productive role of human capital \citep{Lucas1988}. Allowing learning in both sectors makes the opportunity cost of reallocation explicit. We evaluate that cost with education and preferences held fixed. Comparing policies that change preferences raises a further welfare question, considered by \citet{Fabian2025} and discussed in Section~\ref{sec:welfare}.

Section~\ref{sec:model} sets out the economy, and Section~\ref{sec:dynamics} derives its long-run configurations and transitions. Section~\ref{sec:numerical} illustrates the dynamics. Section~\ref{sec:welfare} examines growth and welfare, and Section~\ref{sec:interpretation} returns to the historical question.
\section{Education and the composition of demand}\label{sec:model}
\subsection{Preferences and household demand}
Households consume two bundles. The \emph{basic bundle}, $L$, includes a committed consumption component. The \emph{education-sensitive bundle}, $H$, becomes more valuable to households as education rises. Education is indexed by $E\in\mathcal I$, where $\mathcal I$ is an interval in $[0,\infty)$. The bundles are defined by their role in consumption and can represent different goods or different uses and qualities within an industry.

At positive prices $p_L,p_H$ and income $Y$, a household has preferences represented by
\begin{equation}\label{eq:utility}
 U(C_L,C_H;E)=\left[(1-\alpha(E))^{1/\sigma}(C_L-\gamma(E))^\rho
       +\alpha(E)^{1/\sigma}C_H^\rho\right]^{1/\rho},
 \qquad \rho=\frac{\sigma-1}{\sigma}.
\end{equation}
The utility function has a generalized Stone--Geary form, with $\gamma(E)\geq0$ representing committed consumption of $L$. Here $\sigma$ is the elasticity of substitution between $C_L-\gamma(E)$ and $C_H$, and $\alpha(E)$ is the preference weight on $H$. We initially assume $\sigma>1$, $0<\alpha(E)<1$, and $\alpha'(E)>0$, so education raises the relative preference weight on $H$. On the interior of $\mathcal I$, $\alpha$ is twice continuously differentiable and $\gamma$ is continuously differentiable. Section~\ref{sec:dynamics} considers other values of $\sigma$.

The household chooses $C_L\geq\gamma(E)$ and $C_H\geq0$ subject to $p_LC_L+p_HC_H\leq Y$. We assume $Y>p_L\gamma(E)$, leaving some income for discretionary consumption. We interpret $\gamma(E)$ as committed consumption of $L$, including a fixed physiological minimum. We assume $\gamma'(E)\leq0$ to capture the role of education in reducing avoidable commitments above that minimum.\footnote{Households sometimes treat nonessential purchases as necessities because of limited information or entrenched habits. Other expenditures persist through contractual arrangements, such as automatically renewed subscriptions. The education-dependent component of $\gamma(E)$ captures these avoidable commitments.} The main dynamic analysis starts with $\gamma=0$ and then restores committed consumption.

The preference weight captures what households learn to use or value. Reading offers a concrete example: literacy can raise the value of a book at a given price and income. In this interpretation, $H$ includes activities that draw on reading skills, while committed consumption of $L$ takes priority in the household budget. Education raises the relative value of $H$ through $\alpha(E)$; when $\gamma'(E)<0$, it also leaves more income for discretionary consumption. We hold $E$ fixed within each policy regime.

Let $B=Y-p_L\gamma(E)$ denote discretionary expenditure. Define the relative price $p=p_H/p_L$, the log odds of the preference weight $m(E)=\log[\alpha(E)/(1-\alpha(E))]$, and the logistic function $\Lam(x)=(1+e^{-x})^{-1}$. The fraction of discretionary expenditure devoted to $H$ is
\begin{equation}\label{eq:omega}
 \omega(p,E)=\frac{\alpha(E)p^{1-\sigma}}{1-\alpha(E)+\alpha(E)p^{1-\sigma}}
 =\Lam\bigl(m(E)-(\sigma-1)\log p\bigr).
\end{equation}
The household's demand and total expenditure share are
\begin{align}
 C_H&=\frac{\omega B}{p_H}, &
 C_L&=\gamma(E)+\frac{(1-\omega)B}{p_L},\label{eq:demand}\\
 s_H&\equiv\frac{p_HC_H}{Y}
 =\omega\left(1-\frac{p_L\gamma(E)}{Y}\right).\label{eq:share}
\end{align}
The two shares have different denominators. The discretionary share $\omega$ is measured after committed expenditure; the total expenditure share $s_H$ is measured against all income. Production responds to the latter.

\begin{samepage}
\begin{proposition}[Education, income, and expenditure shares]\label{prop:static}
At fixed prices and income, $\alpha'(E)>0$ and $\gamma'(E)\leq0$ imply $\partial s_H/\partial E>0$. At fixed education and prices, if $\gamma(E)>0$, both goods are normal: the income elasticity of $H$ exceeds one, while that of $L$ lies strictly between zero and one. Education raises the share of an additional unit of income spent on $H$, but the cross derivative $\partial^2s_H/(\partial E\partial Y)$ need not be positive.
\end{proposition}
\end{samepage}

The first result follows from two distinct effects:
\begin{equation}\label{eq:staticE}
 \frac{\partial s_H}{\partial E}
 =\left(1-\frac{p_L\gamma}{Y}\right)\omega(1-\omega)m'(E)
 -\omega\frac{p_L\gamma'(E)}{Y}>0.
\end{equation}
The first term captures the shift in discretionary expenditure toward $H$. The second captures the reduction in avoidable consumption commitments when $\gamma'(E)<0$. Higher income, in turn, raises consumption of both bundles: $\partial C_L/\partial Y=(1-\omega)/p_L>0$ and $\partial C_H/\partial Y=\omega/p_H>0$. When $\gamma>0$, households spend a smaller share on $L$ as income rises, even as they buy more of it.

The fraction of an additional dollar allocated to $H$ is simply $\omega$, whose education derivative is positive. By contrast,
\begin{equation}\label{eq:cross}
 \frac{\partial^2s_H}{\partial E\partial Y}
 =\frac{p_L}{Y^2}\left[\gamma\,\omega(1-\omega)m'(E)+\omega\gamma'(E)\right].
\end{equation}
A lower committed consumption requirement leaves less scope for further income growth to reduce the fraction of the budget absorbed by these commitments. Education can therefore raise the share of $H$ while reducing that share's sensitivity to income.

\subsection{Production, prices, and resource allocation}
Time is discrete, $t=0,1,\ldots$. A unit mass of identical, atomistic households supplies one unit of labor per household. Labor moves freely between competitive sectors. Production is linear in labor:
\begin{equation}\label{eq:production}
 Q_{L,t}=A_{L,t}L_{L,t},\qquad
 Q_{H,t}=A_{H,t}L_{H,t},\qquad L_{L,t}+L_{H,t}=1.
\end{equation}
Households own firms and spend their wage income each period. Constant returns imply zero profits. The economy has full employment and no physical capital, saving, or trade. Given current technology, households choose only current consumption and take the effect of aggregate production on future learning as given.

Normalize the price of $L$ to one. Competitive pricing gives $w_t=A_{L,t}$, $p_t=A_{L,t}/A_{H,t}$, and household income $Y_t=w_t$. We choose fixed reference units for the two bundles and define
\begin{equation}\label{eq:states}
 a_t=\log(A_{H,t}/A_{L,t}),\quad
 r_t=\frac{\gamma(E)}{A_{L,t}},\quad k=\sigma-1>0.
\end{equation}
We refer to $a_t$ as \emph{relative productivity}; it measures the log productivity ratio. The \emph{committed expenditure share}, $r_t$, is the fraction of income needed to buy $\gamma(E)$ units of $L$. To compare productivity across sectors, write $A_{j,t}=\bar A_j\exp(x_{j,t})$, where $\bar A_j>0$ is a fixed technical benchmark and $x_{j,t}$ measures log productivity relative to it. Choose output units such that $\bar A_L=\bar A_H=1$. Then $a_t=x_{H,t}-x_{L,t}$. The catch-up mechanism below compares the two sectors through their productivity indices, $x_{H,t}$ and $x_{L,t}$.\footnote{A change in output units changes $A_j$ and $\bar A_j$ together, leaving $x_j$ unchanged. Transforming the utility representation consistently also preserves expenditure and labor shares.}

Market clearing links expenditure to labor allocation:
\begin{equation}\label{eq:lshare}
 \ell_t\equiv L_{H,t}=s_{H,t}
 =(1-r_t)\Lam\bigl(m(E)+ka_t\bigr).
\end{equation}
Because $p_{H,t}Q_{H,t}=w_tL_{H,t}$ and $Y_t=w_t$, the expenditure share of $H$ equals its labor share, $\ell_t$. More spending on $H$ draws labor into that sector. We use this labor share to measure the activity that generates learning.

\subsection{Learning and technological catch-up}
Under the reference-unit normalization, $x_{j,t}=\log A_{j,t}$. Technology evolves according to
\begin{align}
 x_{H,t+1}-x_{H,t}
 &=g+\nu E+\phi_H\ell_t+\delta\max\{-a_t,0\},\label{eq:techH}\\
 x_{L,t+1}-x_{L,t}
 &=g+\phi_L(1-\ell_t)+\delta\max\{a_t,0\}.\label{eq:techL}
\end{align}
We assume $g>0$, $\nu\geq0$, $\phi_H,\phi_L\geq0$, $\Phi\equiv\phi_H+\phi_L>0$, and $0<\delta\leq1$. Both sectors receive baseline growth $g$. Each also learns from its own activity, at a rate proportional to its labor share. The sector with the lower productivity index benefits from catch-up toward the other sector at rate $\delta$. The leading sector retains its knowledge. The bound $\delta\leq1$ ensures that next period's relative productivity is increasing in its current value.

Learning is external to individual decisions: firms and households take future technology as given when choosing current production and consumption. Education can also raise productivity directly. The coefficient $\nu$ measures the additional effect in $H$ relative to $L$; a common effect in both sectors would cancel from the equation for relative productivity.

Subtracting the technology equations yields
\begin{equation}\label{eq:fullmap}
 a_{t+1}=(1-\delta)a_t+\nu E-\phi_L+\Phi\ell_t.
\end{equation}
Reallocation toward $H$ raises its learning relative to $L$. Catch-up pulls relative productivity toward zero. Relative productivity then affects spending through the relative price in~\eqref{eq:lshare}.

\begin{samepage}
\begin{definition}[Competitive path and balanced growth]\label{def:eq}
Given $E$, initial positive productivities, and $A_{L,0}>\gamma(E)$, a competitive path consists of prices, wages, consumption, employment, and productivities satisfying household optimization, competitive pricing, resource and goods-market clearing, and~\eqref{eq:techH}--\eqref{eq:techL} at every date. When $\gamma=0$, a balanced-growth configuration has constant $a_t=a^*$ and $\ell_t=\ell^*$; the productivity and consumption levels in both sectors then grow at a common constant rate.
\end{definition}
\end{samepage}

The competitive path can be constructed one period at a time. Current technology determines prices and income; household demand determines labor allocation; and learning determines next period's technology. Since $A_{L,t}$ grows at least at log rate $g$, committed consumption remains affordable. Education is exogenous, so changes in $E$ describe changes in consumption capabilities or policy experiments.
\section{Demand feedback and structural transitions}\label{sec:dynamics}
We first set $\gamma=0$, so all consumption is discretionary. Relative productivity alone then determines the allocation dynamics. Equation~\eqref{eq:lshare} gives $\ell_t=\Lam(m(E)+ka_t)$, and the law of motion is
\begin{equation}\label{eq:F}
 a_{t+1}=F(a_t;E)
 =(1-\delta)a_t+\nu E-\phi_L+\Phi\Lam(m(E)+ka_t).
\end{equation}
Higher relative productivity makes $H$ cheaper. With $\sigma>1$, the lower price raises its expenditure share and draws labor into the sector. Learning then raises relative productivity further. Equation~\eqref{eq:F} describes this feedback together with technological catch-up. The same equation will describe the limiting dynamics when the committed expenditure share becomes small.

\subsection{Coexistence of long-run configurations}
Two indices summarize the forces in~\eqref{eq:F}:
\begin{equation}\label{eq:KD}
 K=\frac{\Phi k}{\delta},\qquad
 D(E)=m(E)+\frac{k}{\delta}(\nu E-\phi_L).
\end{equation}
The feedback index $K$ compares the combined effects of substitution and learning with catch-up. The composite index $D(E)$ locates the economy along the set of long-run configurations. It includes both the preference weight and education's direct effect on relative productivity. Since $m'(E)>0$, it rises with education even when $\nu=0$: $D'(E)=m'(E)+k\nu/\delta>0$.

At a balanced-growth configuration, $\delta a^*=\nu E-\phi_L+\Phi\ell^*$. Demand also requires $\logit\ell^*=m(E)+ka^*$, where $\logit\ell\equiv\log[\ell/(1-\ell)]$. Combining these two conditions gives an equation in the labor share alone:
\begin{equation}\label{eq:Z}
 Z(\ell^*)=D(E),\qquad
 Z(\ell)=\log\frac{\ell}{1-\ell}-K\ell,\qquad 0<\ell<1.
\end{equation}
For each root $\ell^*$, relative productivity is $a^*=(\nu E-\phi_L+\Phi\ell^*)/\delta$. Counting the roots of~\eqref{eq:Z} therefore tells us how many balanced-growth configurations exist.

When $K>4$, define
\begin{equation}\label{eq:turning}
 \ell_- =\frac{1-\sqrt{1-4/K}}{2},\qquad
 \ell_+ =\frac{1+\sqrt{1-4/K}}{2},
 \qquad D_-=Z(\ell_+),\quad D_+=Z(\ell_-).
\end{equation}
Here $D_-<D_+$. The subscripts order the shares and composite-index values separately. In particular, the lower value $D_-$ is attained at the larger turning-point share $\ell_+$.

\begin{proposition}[Feedback, multiplicity, and stability]\label{prop:main}
Suppose $\gamma=0$, $\sigma>1$, and the technology restrictions in Section~\ref{sec:model} hold.
\begin{enumerate}
\item If $K\leq4$, there is a unique balanced-growth configuration, and every finite initial $a_0$ converges to it. If $K<4$, the map is a global contraction.
\item If $K>4$ and $D_-<D(E)<D_+$, there are exactly three configurations, ordered by their shares and relative productivity. The low-share and high-share configurations are locally asymptotically stable; the middle configuration is unstable.
\item In the three-root case, every initial state below the middle relative-productivity root converges to the low-share configuration, and every initial state above it converges to the high-share configuration. A path beginning exactly at the middle root remains there.
\item If $K>4$ and $D(E)$ lies outside $[D_-,D_+]$, there is exactly one configuration, which attracts every initial state. At either endpoint, there are two distinct configurations, one of which is a double root and is stable from only one side.
\end{enumerate}
\end{proposition}

The shape of $Z$ determines the number of configurations. Its derivative is $Z'(\ell)=1/[\ell(1-\ell)]-K$. Since $\ell(1-\ell)\leq1/4$, $Z$ is increasing when $K\leq4$ and crosses each value of $D$ once. When $K>4$, it has a local maximum and a local minimum. A horizontal line between them crosses $Z$ three times.

Stability depends on how next period's relative productivity responds to the current state. At a fixed point, that response is
\begin{equation}\label{eq:multiplier}
 F_a(a^*;E)=1-\delta+\Phi k\ell^*(1-\ell^*)
 =1-\delta\{1-K\ell^*(1-\ell^*)\}.
\end{equation}
This multiplier is nonnegative. At the two stable configurations, it is below one, so small departures shrink over time. At the middle configuration, it exceeds one, so departures grow. Moreover, the map preserves the ordering of states: a path cannot jump across a fixed point. This property makes the middle configuration the boundary between paths that converge to the low-share and high-share configurations. Appendix~\ref{app:proofs} proves the global results and treats the boundary case $K=4$.

Coexistence thus requires a strong enough expenditure response to prices and a strong enough learning response to labor reallocation. Catch-up must be weak enough for these two responses to reinforce a productivity advantage over some range.

\subsection{Education and structural transitions}
As education rises, both the share of $H$ and its relative productivity increase along a stable branch. Write that branch as $\ell^*(E),a^*(E)$. A \emph{fold} is a point where a stable branch meets an unstable one. Such a fold is attainable if the corresponding education level lies in the interior of $\mathcal I$ and $D(E)$ reaches $D_-$ or $D_+$. A bounded preference shift may leave either threshold out of reach.

\begin{proposition}[Education and the transition threshold]\label{prop:education}
At a hyperbolically stable configuration,
\begin{align}
 \frac{\dd\ell^*}{\dd E}
 &=\frac{\ell^*(1-\ell^*)D'(E)}{1-K\ell^*(1-\ell^*)}>0,\label{eq:dell}\\
 \frac{\dd a^*}{\dd E}
 &=\frac{\nu+\Phi\ell^*(1-\ell^*)m'(E)}{\delta-\Phi k\ell^*(1-\ell^*)}>0.\label{eq:da}
\end{align}
Consequently $p^*=e^{-a^*}$ decreases. If $K>4$ and both folds are attainable, their education levels are uniquely defined by
\begin{equation}\label{eq:thresholds}
 D(E_-)=D_-,\qquad D(E_+)=D_+,\qquad E_-<E_+.
\end{equation}
As education rises through $E_+$, the low-share and middle configurations disappear in a nondegenerate saddle-node bifurcation. At any fixed $E>E_+$ in the admissible interval, the remaining high-share configuration attracts every initial state. These conclusions also hold when $\nu=0$.
\end{proposition}

Education shifts expenditure toward $H$, and learning amplifies that shift through lower relative prices. The positive denominators in~\eqref{eq:dell}--\eqref{eq:da} measure how much catch-up restrains this amplification on a stable branch. The demand channel is sufficient for the result: setting $\nu=0$ removes education's direct productivity effect but leaves the feedback intact. A positive $\nu$ changes the education thresholds through $D(E)$.

The unstable branch moves in the opposite direction: its share falls as education rises. At a fold, a stable branch meets this boundary and ends. Near that point, a small education change has a large long-run effect, but adjustment is slow because the convergence multiplier is close to one.

A temporary increase in $E$ can also leave a lasting effect. It must last long enough to move relative productivity across the original boundary before $E$ returns to its starting value.

\begin{corollary}[Persistence after temporary support]\label{cor:pulse}
Let $E_0$ be in the three-root region, with relative-productivity roots $a_1^*<a_2^*<a_3^*$. Start at $a_1^*$, and temporarily apply an attainable $E_1>E_+$. There is a finite duration after which the state exceeds $a_2^*$. If education then returns to $E_0$, the economy converges to $a_3^*$. If support ends with the state strictly below $a_2^*$, it returns to $a_1^*$.
\end{corollary}

For this experiment, $E$ measures effective education exposure or support for education-related consumption. The temporary increase shifts spending and changes relative productivity through learning. Its lasting effect depends on the productivity state reached when the intervention ends. This interpretation treats the temporary change in $E$ as exposure or support, rather than as a loss of completed schooling when it ends.

\subsection{The role of substitution and catch-up}
In the Cobb--Douglas limit, discretionary spending satisfies $\omega=\alpha(E)$ independently of prices. With $\gamma=0$,
\begin{equation}\label{eq:CD}
 a_{t+1}=(1-\delta)a_t+\nu E-\phi_L+\Phi\alpha(E),\qquad
 a^*=\frac{\nu E-\phi_L+\Phi\alpha(E)}{\delta}.
\end{equation}
The unique configuration is globally stable. Education still shifts expenditure and learning, but prices no longer amplify that shift. This is also the discretionary allocation in the Stone--Geary specification.

When $0<\sigma<1$, a lower price for $H$ reduces its expenditure share. This negative feedback gives a unique balanced-growth configuration. It may nevertheless produce oscillations in discrete time. A local multiplier between minus one and one ensures stability; a value below minus one implies instability. At minus one, nonlinear terms determine stability. Appendix~\ref{app:benchmarks} gives the condition in terms of the model parameters.

Removing catch-up has a different effect. With $\delta=0$ and $\sigma>1$, any finite fixed point is unstable: a small productivity advantage reinforces itself. Catch-up is therefore the force that allows the stable outer configurations in the main model.

\subsection{Reintroducing committed consumption}\label{sec:subsistence}
With $\gamma(E)>0$, relative productivity and the committed expenditure share jointly determine the path. Holding education fixed, their laws of motion are
\begin{align}
 a_{t+1}&=F(a_t;E)-\Phi r_t\Lam(m(E)+ka_t),\label{eq:perturb}\\
 r_{t+1}&=r_t\exp\{-g-\phi_L(1-\ell_t)-\delta\max(a_t,0)\}.\label{eq:rmap}
\end{align}
Feasibility requires $0\leq r_0<1$. Growth reduces this share at least geometrically: $r_t\leq r_0e^{-gt}$. Producing the fixed quantity of committed consumption therefore absorbs a shrinking fraction of labor. The question is whether that early use of resources still affects the long-run allocation.

\begin{proposition}[A vanishing committed expenditure share]\label{prop:subsistence}
If $K<4$, every feasible path of~\eqref{eq:perturb}--\eqref{eq:rmap} converges in relative productivity and expenditure share to the unique configuration characterized in Proposition~\ref{prop:main}. If $K>4$, the limiting fixed-point equation remains~\eqref{eq:Z}. In the three-root region, $D_-<D(E)<D_+$, a positive initial committed expenditure share can change the eventual branch relative to the economy with $r_0=0$.
\end{proposition}

Under weak feedback, the economy eventually reaches the same configuration as an economy without consumption commitments. Under strong feedback, early spending to meet these commitments can change the destination. It directs labor and learning toward $L$, possibly moving relative productivity below the boundary that separates the two stable configurations in the limiting system. The committed expenditure share then vanishes, but the accumulated difference in productivity remains relevant. During the transition, the destination depends on both relative productivity and the committed expenditure share.
\section{Illustrative dynamics}\label{sec:numerical}
The examples show how the economy moves between configurations and why its starting point matters. We use the parameters in Table~\ref{tab:parameters} and set $\alpha(E)=\Lam(m_0+\eta E)$, so that $m(E)=m_0+\eta E$. These are illustrative values; education and time are measured in model units.

\begin{table}[H]
\centering
\caption{Parameters of the illustrative economy}\label{tab:parameters}
\small\begin{tabular}{llr}
\toprule
Parameter & Interpretation & Value \\
\midrule
$\delta$ & Technology catch-up rate & 0.05 \\
$\phi_H$ & H-sector learning coefficient & 0.12 \\
$\phi_L$ & L-sector learning coefficient & 0.08 \\
$\sigma$ & Elasticity of substitution & 3 \\
$g$ & Common productivity drift & 0.005 \\
$\nu$ & Direct education effect & 0.01 \\
$\eta$ & Preference response to education & 1 \\
$m_0$ & Preference intercept & -2.2 \\
\bottomrule
\end{tabular}
\figurenote{The baseline sets $\gamma=0$. Education $E$ measures effective learning exposure. Figure~\ref{fig:extensions} compares weaker substitution and an initial committed expenditure share.}
\end{table}

\subsection{Education and the long-run allocation}
\begin{figure}[!htbp]
\centering
\includegraphics[width=0.94\textwidth]{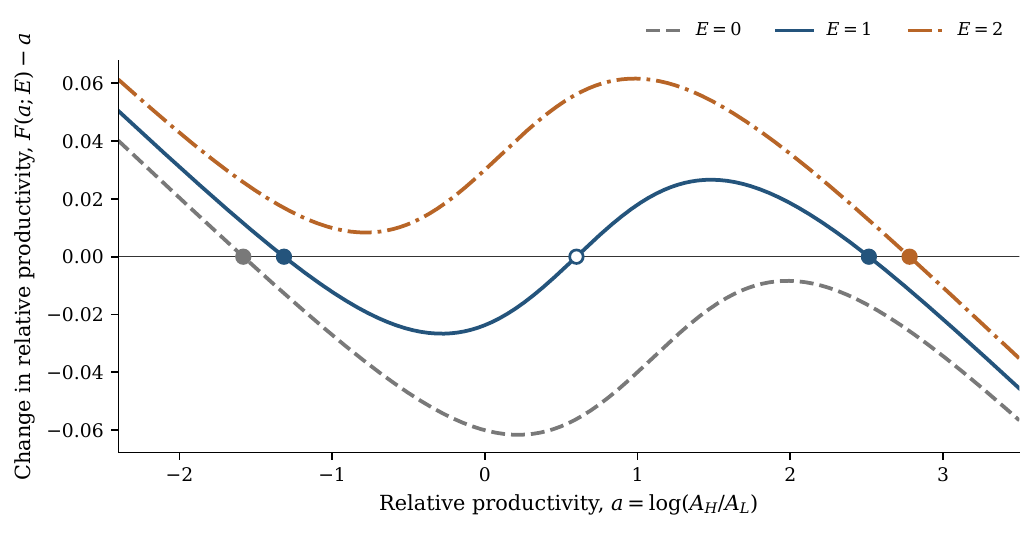}
\caption{Education and the change in relative productivity}\label{fig:phase}
\figurenote{The three curves plot $F(a;E)-a$ under the parameters in Table~\ref{tab:parameters}. Positive drift raises $a$ and negative drift lowers it. The $E=1$ curve has two stable zeros separated by an unstable zero. Each other curve has one globally attracting zero. Filled markers indicate stable roots and the open marker an unstable root.}
\end{figure}

The baseline has strong feedback: $K=8$. Its composite index is $D(E)=-5.4+1.4E$, and the two education thresholds are $E_-=0.238800$ and $E_+=1.761200$. Figure~\ref{fig:phase} compares education levels below, between, and above these thresholds. Each curve shows the change in relative productivity from one period to the next. A zero is a balanced-growth configuration; the sign of the curve shows which way the economy moves.

At $E=1$, both a small and a large sector $H$ can persist. Its two stable expenditure shares are about $0.02125$ and $0.97875$. The unstable share between them is $0.5$, corresponding to relative productivity $a_2^*=0.6$. Starting below that productivity threshold leads to the low-share configuration; starting above it leads to the high-share configuration. Table~\ref{tab:roots} reports these roots and their local multipliers. At $E=0$ and $E=2$, the economy has a unique, globally stable configuration.

\begin{table}[H]
\centering
\caption{Balanced-growth configurations and local stability}\label{tab:roots}
\small\begin{tabular}{rrrrl}
\toprule
$E$ & $\ell^*$ & $a^*$ & $F_a(a^*;E)$ & Stability \\
\midrule
0.0 & 0.004667 & -1.581334 & 0.951858 & Stable \\
1.0 & 0.021248 & -1.315008 & 0.958319 & Stable \\
1.0 & 0.500000 & 0.600000 & 1.050000 & Unstable \\
1.0 & 0.978752 & 2.515008 & 0.958319 & Stable \\
2.0 & 0.995333 & 2.781334 & 0.951858 & Stable \\
\bottomrule
\end{tabular}
\figurenote{The multiplier measures how next period's relative productivity responds to the current state. A value strictly between zero and one implies local asymptotic stability. The relative price is $p^*=\exp(-a^*)$.}
\end{table}

\begin{figure}[H]
\centering
\includegraphics[width=0.88\textwidth]{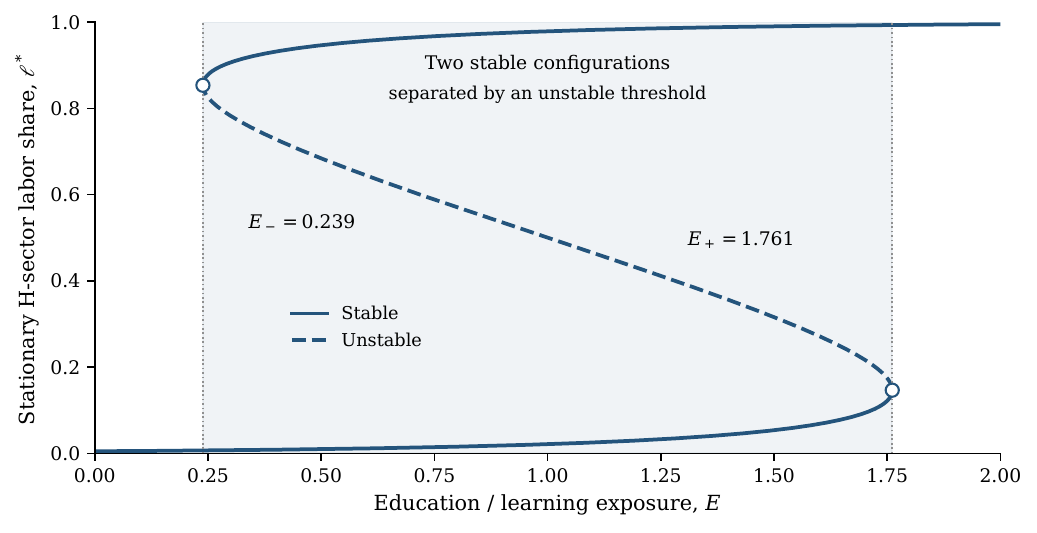}
\caption{Balanced-growth expenditure shares and education}\label{fig:bifurcation}
\figurenote{Away from the folds, solid branches are locally stable and the dashed branch is unstable. Open circles mark the fold points, where the local multiplier equals one and each configuration attracts from only one side. The lower and upper education thresholds are $E_-=0.238800$ and $E_+=1.761200$. Increasing education beyond the upper threshold eliminates the low-share configuration. Decreasing education below the lower threshold eliminates the high-share configuration.}
\end{figure}

Figure~\ref{fig:bifurcation} traces the configurations as education changes. The two stable branches coexist between $E_-$ and $E_+$.

Education can produce the same type of transition through demand alone. Setting $\nu=0$ leaves $K=8$ and gives $D(E)=-5.4+E$. The thresholds become $E_-=0.334320$ and $E_+=2.465680$. Removing the direct productivity effect thus raises the education level needed to end the low-share branch. In this example, that level lies beyond the displayed interval $[0,2]$.

\subsection{The duration of temporary exposure}
\begin{figure}[H]
\centering
\includegraphics[width=0.85\textwidth]{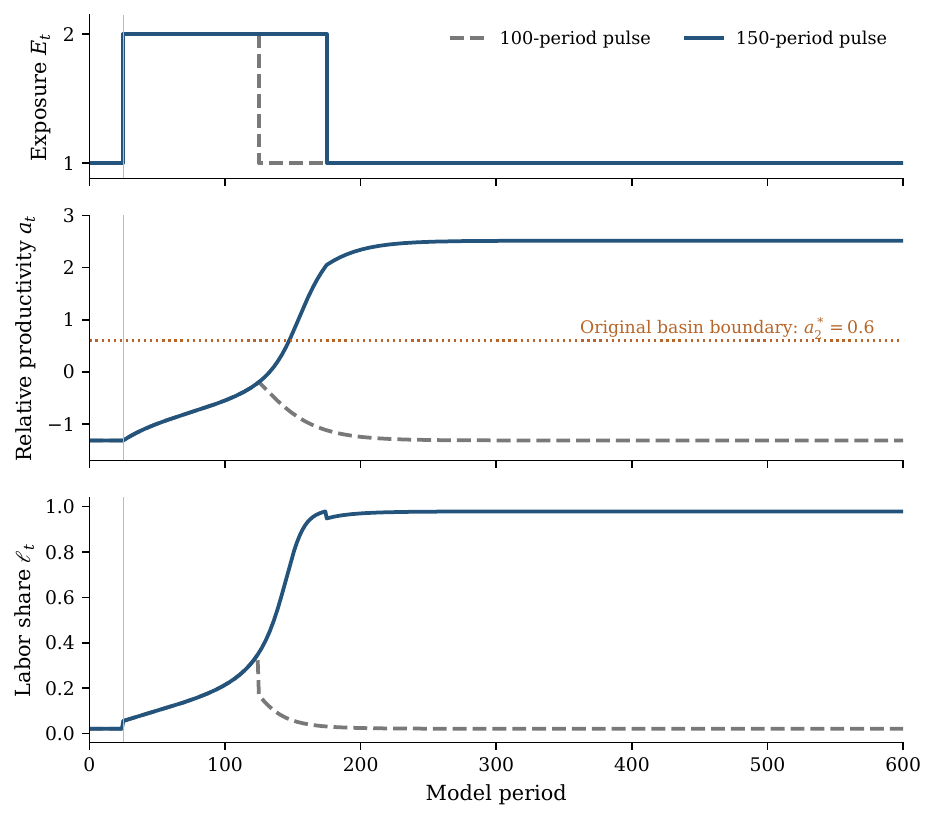}
\caption{Temporary learning exposure and persistent reallocation}\label{fig:pulse}
\figurenote{Both paths begin at the same low-share configuration and return to $E=1$. The 100-period increase lasts from $t=25$ through $t=124$; the 150-period increase lasts through $t=174$. Only the longer path crosses the original productivity threshold $a_2^*=0.6$. Expenditure shares can jump when exposure changes; productivity adjusts through subsequent learning.}
\end{figure}

Figure~\ref{fig:pulse} starts the economy at the low-share configuration for $E=1$. At period 25, education exposure rises to $E=2$. The two paths differ only in how long this increase lasts. After 100 periods, relative productivity is $-0.1961$, still below the original threshold of $0.6$. Returning exposure to $E=1$ then sends the economy back to the low-share configuration. After 150 periods, relative productivity has reached $2.0515$. The economy now approaches the high-share configuration even after exposure returns to $E=1$. For these parameters, the shortest duration that crosses the original threshold is 123 periods.

\subsection{Substitution and early consumption commitments}
\begin{figure}[H]
\centering
\includegraphics[width=\textwidth]{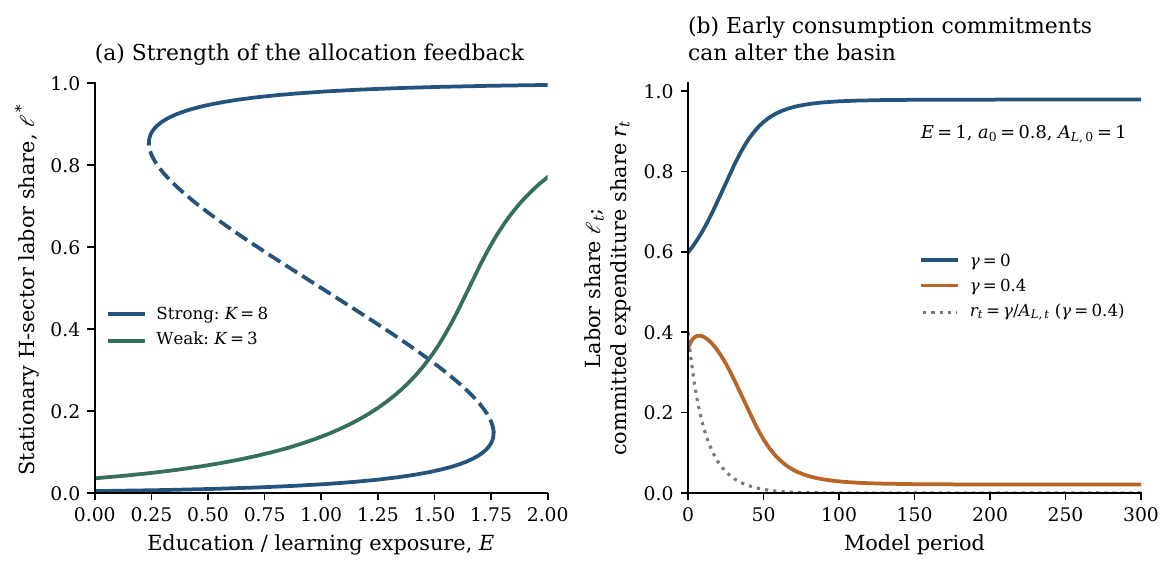}
\caption{Weaker substitution and an initial committed expenditure share}\label{fig:extensions}
\figurenote{The left panel changes only $\sigma$, giving $K=3$. The right panel uses the baseline $K=8$ and identical initial productivities, changing $\gamma$ from zero to $0.4$ and hence $r_0$ from zero to $0.4$. The labor share of $H$ is $\ell_t=(1-r_t)\omega_t$ in both technology equations.}
\end{figure}

The left panel of Figure~\ref{fig:extensions} lowers the substitution elasticity from $3$ to $1.75$. Feedback falls from $K=8$ to $K=3$, and only one configuration remains at each education level. The learning coefficients are unchanged; the difference comes from the weaker expenditure response to prices.

The right panel restores the baseline parameters and compares two economies with $E=1$, $a_0=0.8$, and $A_{L,0}=1$. Without consumption commitments, the economy approaches the high-share configuration. With $\gamma=0.4$, early spending to meet these commitments draws enough activity toward $L$ to push relative productivity below $0.6$ in period 7. The economy then approaches the low-share configuration. Committed consumption stays fixed, while its expenditure share $r_t=\gamma/A_{L,t}$ vanishes as productivity rises. Its lasting effect comes from the learning accumulated during the early transition.

Appendix~\ref{app:numerics} varies each reported parameter by ten percent in either direction, holding the others fixed. All of these changes preserve $K>4$ and three configurations at $E=1$. The thresholds move, and some lower folds fall below zero, outside the admissible education domain.
\section{Growth and the social value of reallocation}\label{sec:welfare}
Reallocation toward $H$ affects future productivity through learning gained in $H$ and forgone in $L$. We first compare growth across balanced-growth configurations and then examine the social value of this reallocation. Throughout this section, we set $\gamma=0$.

\subsection{Balanced-growth rates}
Hold education fixed. Each sector's productivity growth has three components: baseline growth, its own learning, and catch-up. Write the contributions of education and own-sector learning as
\begin{equation}\label{eq:bases}
 b_H=\nu E+\phi_H\ell^*,\qquad b_L=\phi_L(1-\ell^*).
\end{equation}
The sector with the larger contribution leads in relative productivity. At a balanced-growth configuration, the productivity gap satisfies $\delta a^*=b_H-b_L$. Catch-up closes the difference between the sectors' growth rates.

\begin{proposition}[Common growth and its ranking]\label{prop:growth}
Both productivities and both consumption quantities grow at the common log rate
\begin{equation}\label{eq:growth}
 G^*=g+\max\{\nu E+\phi_H\ell^*,\;\phi_L(1-\ell^*)\}.
\end{equation}
With fixed education and preference weights, the consumption aggregate in~\eqref{eq:utility} grows at the same rate. At fixed $E$, its growth rate increases with $\ell^*$ in the region $a^*>0$ when $\phi_H>0$, and decreases with $\ell^*$ in the region $a^*<0$ when $\phi_L>0$. A high-share configuration need not have a higher common growth rate than a low-share configuration.
\end{proposition}

The common growth rate is set by the leading sector. If $H$ leads, it grows at $g+b_H$, and catch-up allows $L$ to grow at the same rate. If $L$ leads, both sectors grow at $g+b_L$. Constant labor shares then give both consumption quantities the same growth rate. The consumption aggregate inherits that rate because it is homogeneous of degree one.

Let $G_1^*$ and $G_3^*$ be the common log growth rates at two coexisting configurations with shares $\ell_1^*<\ell_3^*$. Their difference is
\begin{equation}\label{eq:growthcomparison}
 G_3^*-G_1^*=
 \max\{\nu E+\phi_H\ell_3^*,\phi_L(1-\ell_3^*)\}
 -\max\{\nu E+\phi_H\ell_1^*,\phi_L(1-\ell_1^*)\}.
\end{equation}
Suppose $L$ leads at the low-share configuration and $H$ leads at the high-share configuration. The latter grows faster exactly when $\nu E+\phi_H\ell_3^*>\phi_L(1-\ell_1^*)$. The numerical baseline satisfies this inequality. Appendix~\ref{app:planner} gives an example in which the inequality is reversed. The comparison depends on how much each sector learns, whereas the feedback index $K$ uses the sum of the two learning coefficients.

This distinction also matters along a stable branch. While $L$ remains the leader, education draws labor away from the sector that sets the common growth rate. Since $G^*=g+\phi_L(1-\ell^*)$, we have $\dd G^*/\dd E=-\phi_L\,\dd\ell^*/\dd E<0$ when $\phi_L>0$. Education can therefore raise the share of $H$ while lowering long-run growth locally. To assess welfare, we must also account for what households consume along the transition.

\subsection{The learning gained and forgone}
Household spending determines sectoral labor allocation in equilibrium. Individual households do not internalize the resulting effects on future productivity. A planner takes these effects into account. To compare their choices under the same preferences, hold $E$ fixed and consider dates $t=0,\ldots,T$. The planner chooses the labor share $\ell_t\in[0,1]$, and households consume what the two sectors produce:
\begin{equation}\label{eq:plannerresources}
 C_{L,t}=e^{x_{L,t}}(1-\ell_t),\qquad C_{H,t}=e^{x_{H,t}}\ell_t.
\end{equation}
The planner maximizes $\sum_{t=0}^{T}\beta^t u(x_{L,t},x_{H,t},\ell_t;E)$, where $\beta\in(0,1)$ is the discount factor and $u$ is log utility from the consumption quantities in~\eqref{eq:plannerresources}. Technology follows~\eqref{eq:techH}--\eqref{eq:techL} with the chosen labor share.

Let $V_t(x_L,x_H;E)$ be the value of the planner's problem from date $t$, with $V_{T+1}=0$. Its derivative $V_{j,t+1}$ is the marginal value of next period's log productivity in sector $j$, evaluated at the chosen next state. At a differentiable interior optimum,
\begin{equation}\label{eq:plannerFOC}
 u_{\ell,t}
 +\beta\left[\phi_H V_{H,t+1}-\phi_L V_{L,t+1}\right]=0.
\end{equation}
The first term is the effect of reallocation on current utility. The bracketed term values the extra learning in $H$ and subtracts the value of learning lost in $L$. Current relative productivity fixes the catch-up terms, so reallocating labor affects next period's technology directly through these two learning terms. The continuation values also account for subsequent catch-up.

\begin{samepage}
\begin{proposition}[The net learning wedge]\label{prop:wedge}
Suppose $V_{t+1}$ is differentiable at the next state induced by the decentralized current allocation. That allocation maximizes current utility and satisfies $u_{\ell,t}=0$. The planner's additional marginal return to reallocating labor toward $H$ is
\begin{equation}\label{eq:wedge}
 \mathcal W_t=\beta\left[\phi_H V_{H,t+1}-\phi_L V_{L,t+1}\right].
\end{equation}
Evaluated at the decentralized current allocation, $\mathcal W_t>0$ means that a small shift of labor toward $H$ raises the planner's objective when later allocations are chosen optimally. If $\mathcal W_t<0$, a small shift toward $L$ raises it. The sign depends on the value of learning in both sectors.
\end{proposition}
\end{samepage}

We call $\mathcal W_t$ the \emph{net learning wedge}. At the competitive allocation, the marginal gain in current utility is zero. The wedge therefore identifies whether a small shift toward $H$ would raise or lower the planner's objective, with later allocations chosen optimally. It is a local comparison and remains valid when that objective is nonconcave. At date $T$, future learning has no value within the horizon, and the wedge is zero. Appendix~\ref{app:planner} derives the first-order and continuation-value equations.

\subsection{Education and the value of learning}
Education can change learning even when its direct productivity effect is zero. At a fixed current state, competitive households respond to education by raising the labor share of $H$:
\begin{equation}\label{eq:lE}
 \ell_E=\ell(1-\ell)m'(E)>0.
\end{equation}
This reallocation changes learning in both sectors. Education may also raise productivity in $H$ directly through $\nu$. Under the differentiability condition in Proposition~\ref{prop:wedge}, the value of these two effects on next period's technology is
\begin{equation}\label{eq:educationexternal}
 \beta\nu V_{H,t+1}
 +\beta\ell_E\left[\phi_H V_{H,t+1}-\phi_L V_{L,t+1}\right].
\end{equation}
The first term values the direct productivity effect. The second equals $\ell_E\mathcal W_t$: it values the demand shift at the net learning wedge. With $\nu=0$, education raises the value of next period's technology precisely when the learning gained in $H$ is worth more than the learning forgone in $L$. Evaluating an education policy also requires its current consumption effects, its costs, and the effects of any change in future education.

Because education changes preferences, policy evaluation needs a common welfare criterion. For a fixed $E$, multiplying utility by any positive factor $\chi(E)$ preserves household choices. Across education levels, however, the same transformation adds $\log\chi(E)$ to log utility and can change welfare rankings. One approach is to evaluate all consumption paths using a fixed reference preference index. Other approaches ask which choices or preferences should guide the comparison: \citet{Bernheim2009} use unambiguous choice, while \citet{Fabian2025} examine preferences formed through learning and reflection. Under the chosen welfare criterion, the demand channel's contribution to future technology is valued by $\ell_E\mathcal W_t$.
\section{Consumption capabilities in historical perspective}\label{sec:interpretation}
How did households come to value particular goods, and how did their purchases expand markets? De Vries's account of the industrious revolution gives household demand a role in the shift toward market work \citep{DeVries1994}. Our model links that demand to consumption capabilities. Learning to use a good can change its place in household spending and redirect productive activity.

Books make this question concrete. Reading skills, the cost of a book, and the income left after necessities all affect whether a household buys one. The historical expansion described by \citet{Buringh2009} brings these forces together. The model helps separate their roles. A household may buy more books because it has more to spend, because books have become cheaper, or because reading has become more valuable to it. Each explanation points to a different part of the historical record.

Household budgets are especially useful here. Purchases recorded alongside expenditure on necessities can show whether a growing market reflects rising disposable resources or a change in how those resources are used. Literacy and schooling records can help describe the skills of potential readers. Price and purchase records together would show whether cheaper books increased their expenditure share, as required by the feedback mechanism. Bringing these records together would give the demand channel a clearer empirical meaning than a correlation between schooling and total purchases alone.

The next historical question concerns producers. An expansion of purchases may draw workers into the activities that supply the new demand. The mechanism studied here requires that this activity improve productivity and that lower relative prices sustain the expenditure shift. Production, employment, and price records would therefore be needed alongside household evidence. Their timing matters: the argument concerns a sequence from changing consumption to productive activity and then to the prices faced by later consumers. Establishing the direction of these links requires variation that separates changes in consumer capabilities from simultaneous changes in production technology.

Comparisons across places would also need to recover their starting points. Similar literacy rates at a later date can conceal different histories of purchasing power, basic needs, and productive experience. A region in which households could support an expanding market earlier may have accumulated more experience in supplying it. The relevant comparison would follow those histories, rather than explain later production patterns using current schooling alone. This is where household consumption can make a distinct contribution to the study of persistent differences between economies.

The model treats these changes as a reallocation of fully employed labor. Historical episodes with substantial idle resources raise an additional question about the level of activity. Effective-demand failures of the kind studied by \citet{Woodford2022} provide a way to approach that question. Combining the two perspectives would allow changes in what households buy to affect both which activities expand and how much labor is employed.

A quantitative application would require estimates of how education shifts demand, how expenditure responds to relative prices, and how sectoral activity affects subsequent productivity. Together with the rate of technological catch-up, these estimates would determine whether the feedback is strong enough to sustain multiple configurations.

\section{Conclusion}
Education can change production through consumption. In the model, a shift in household demand reallocates labor and learning. When substitution and learning are strong relative to catch-up, that shift can sustain a different long-run configuration.

Early consumption commitments can influence which activities accumulate productive experience and hence where the economy settles. The growth and welfare consequences depend on learning gained and forgone across sectors. What households learn to use can therefore influence what an economy learns to produce.
\clearpage
\begin{singlespace}
\small

\end{singlespace}
\clearpage
\appendix
\section{Household demand and static comparisons}\label{app:household}
Write $X_L=C_L-\gamma(E)$ and $X_H=C_H$. At fixed $E$, the household allocates $B=Y-p_L\gamma(E)>0$ to a homothetic CES aggregate. Both quantities are positive at the optimum, because the marginal utility of a missing bundle is unbounded when $\sigma>1$. The ratio of marginal utilities gives
\begin{equation}
 \frac{X_H}{X_L}=\frac{\alpha}{1-\alpha}\left(\frac{p_H}{p_L}\right)^{-\sigma}.
\end{equation}
Multiplying by the price ratio yields the expenditure odds in~\eqref{eq:omega}; using the budget constraint gives~\eqref{eq:demand}. The expenditure index for one unit of the CES aggregate is
\begin{equation}
 P(p_L,p_H;E)=\left[(1-\alpha(E))p_L^{1-\sigma}
                      +\alpha(E)p_H^{1-\sigma}\right]^{1/(1-\sigma)}.
\end{equation}
Indirect utility is $B/P$ at fixed $E$, with the CES price index converting discretionary expenditure into utility.

\begin{proof}[Proof of Proposition~\ref{prop:static}]
At fixed prices, differentiating the logistic expression gives $\omega_E=\omega(1-\omega)m'(E)>0$. Equation~\eqref{eq:staticE} follows by differentiating~\eqref{eq:share}; its first term is strictly positive and its second term is nonnegative. Holding education fixed, the income elasticities are
\begin{equation}
 \varepsilon_H=\frac{Y}{Y-p_L\gamma}>1,\qquad
 \varepsilon_L=\frac{(1-\omega)Y}{p_L\gamma+(1-\omega)(Y-p_L\gamma)}\in(0,1),
\end{equation}
where the strict inequalities use $\gamma>0$. The quantity derivatives are positive. The expenditure derivative satisfies $\partial(p_HC_H)/\partial Y=\omega$, whose education derivative is positive. Differentiating $\partial s_H/\partial Y=\omega p_L\gamma/Y^2$ yields~\eqref{eq:cross}. Its two terms may have opposite signs. If $\gamma=0$, both income elasticities equal one and the total share is independent of income.
\end{proof}

For market clearing, competitive firms satisfy $p_jA_j=w$. With labor supply one and zero profits, $Y=w$. Hence $L_H=p_HC_H/w=s_H$. Under the normalization $p_L=1$, $w=A_L$ and $p_H=A_L/A_H$, producing~\eqref{eq:lshare}. The goods quantities from~\eqref{eq:demand} equal $A_LL_L$ and $A_HL_H$, so the construction clears both goods and labor markets.

\section{Proofs for relative dynamics}\label{app:proofs}
The proofs use relative productivity $a=x_H-x_L$. Define its change per period as $f(a;E)=F(a;E)-a$. At fixed education and with $\gamma=0$, write the common labor and expenditure share as $q(a)=\Lam(m(E)+ka)$. Two identities organize the argument:
\begin{align}
 q_a&=kq(1-q),\qquad
 f_a=-\delta+\Phi kq(1-q),\label{eq:proofderiv}\\
 f(a;E)&=\frac{\delta}{k}\{D(E)-Z(q(a))\}.\label{eq:proofdrift}
\end{align}
The map is strictly increasing for finite $a$ because $F_a=1-\delta+\Phi kq(1-q)>0$. Every orbit is bounded: $F(a)$ lies between $(1-\delta)a+c_L$ and $(1-\delta)a+c_H$, where $c_L=\nu E-\phi_L$ and $c_H=\nu E+\phi_H$. Thus
\begin{equation}\label{eq:bound}
 \min\{a_0,c_L/\delta\}\leq a_t\leq\max\{a_0,c_H/\delta\}
 \quad\text{for all }t.
\end{equation}

\begin{proof}[Proof of Proposition~\ref{prop:main}]
\textit{Existence and root count.} The function $Z$ tends to minus infinity as $\ell\downarrow0$ and to plus infinity as $\ell\uparrow1$. Its derivative is $1/[\ell(1-\ell)]-K$. If $K<4$ it is strictly positive. If $K=4$ it is nonnegative and vanishes only at $1/2$; integrating its derivative over any nonempty interval still gives a strictly positive increase. Thus there is exactly one root for $K\leq4$.

If $K>4$, the derivative vanishes precisely at $\ell_-$ and $\ell_+$ in~\eqref{eq:turning}. It is positive on $(0,\ell_-)$, negative on $(\ell_-,\ell_+)$, and positive on $(\ell_+,1)$. Consequently there are three roots for $D_-<D<D_+$, one for $D<D_-$ or $D>D_+$, and two distinct roots at either endpoint. Since $Z''(\ell)=-(1-2\ell)/[\ell^2(1-\ell)^2]$, the tangency at each endpoint is a double root for $K>4$.

\textit{Local stability.} Equation~\eqref{eq:multiplier} follows from~\eqref{eq:proofderiv}. At a root with $Z'>0$, $K\ell(1-\ell)<1$, giving $0<F_a<1$. At a root with $Z'<0$, $F_a>1$. A double root has $F_a=1$ and nonzero $f_{aa}$, producing attraction on one side and repulsion on the other.

\textit{Global dynamics.} The map preserves order. In particular, if $a<a^*$ for a fixed point, then $F(a)<a^*$, and an orbit cannot cross that fixed point in one step. In the unique-root cases, $f$ is positive below the root and negative above it, because it is continuous and has no other zero and its tails have opposite signs. Starting on either side, the orbit moves monotonically toward the root without crossing it. Boundedness and continuity imply convergence to that root.

In the three-root case, the sign of $f$ from left to right is $+,-,+,-$. Below the middle root, the orbit either rises toward the low root or falls toward it; it cannot cross either bounding fixed point. The analogous argument gives convergence to the high root above the middle root. The middle root itself is stationary. At a fold the same drift argument establishes the one-sided stability of the double root.

Finally, for $K<4$, the uniform derivative bound
\begin{equation}
 0\leq F_a(a;E)\leq1-\delta+\delta K/4<1
\end{equation}
proves global contraction. At $K=4,D=-2$, the unique root has $\ell=1/2$ and multiplier one; the global drift argument establishes convergence at this nonhyperbolic root.
\end{proof}

\begin{proof}[Proof of Proposition~\ref{prop:education}]
At a simple root, implicit differentiation of $Z(\ell^*)=D(E)$ gives $\dd\ell^*/\dd E=D'/Z'$, which is~\eqref{eq:dell}. Differentiating $f(a^*(E);E)=0$ gives
\begin{equation}
 \frac{\dd a^*}{\dd E}
 =\frac{\nu+\Phi q(1-q)m'}{\delta-\Phi kq(1-q)}.
\end{equation}
The denominator is positive on stable branches. The numerator is strictly positive because $m'>0$, $\Phi>0$, and $q\in(0,1)$. The price derivative is $\dd p^*/\dd E=-p^*\dd a^*/\dd E$.

For $K>4$, let $(a_f,E_f)$ denote a point at either fold. The fold conditions $f=0$ and $f_a=0$ hold, while
\begin{equation}
 f_{aa}=\Phi k^2q(1-q)(1-2q)\ne0,\qquad
 f_E=\nu+\Phi q(1-q)m'>0.
\end{equation}
The inequalities use $q=\ell_-$ or $\ell_+$, neither of which is $1/2$. These are the curvature and transversality conditions for a nondegenerate saddle-node of a scalar map. Because $D(E)$ increases strictly, the folds are crossed in the stated order whenever they lie in the attainable education interval. The root classification in Proposition~\ref{prop:main} identifies which branches disappear. None of this requires $\nu>0$.
\end{proof}

\begin{proof}[Proof of Corollary~\ref{cor:pulse}]
For all fixed $a$, $F(a;E)$ increases strictly with $E$. At $E_1>E_+$, its unique fixed point $a^*(E_1)$ lies above $a_2^*$, since $F(a_2^*;E_1)>F(a_2^*;E_0)=a_2^*$. Starting at $a_1^*$, the $E_1$ orbit increases monotonically to $a^*(E_1)$. Since that limit is strictly above $a_2^*$, the orbit crosses $a_2^*$ after finitely many steps. On returning to $E_0$, Proposition~\ref{prop:main} determines the destination from the state at the return date. Exact equality with $a_2^*$ is the knife-edge case in which the path remains at the unstable configuration.
\end{proof}

\section{Committed consumption and benchmark cases}\label{app:benchmarks}
\begin{proof}[Proof of Proposition~\ref{prop:subsistence}]
For a feasible initial condition, $0\leq r_t\leq r_0e^{-gt}<1$. Since $0\leq\ell_t\leq1$, the bounds~\eqref{eq:bound} continue to apply. When $K<4$, let $\kappa=1-\delta+\delta K/4<1$ be a Lipschitz constant for $F$, and let $a^*$ denote its unique fixed point. Equation~\eqref{eq:perturb} gives
\begin{equation}
 |a_{t+1}-a^*|\leq \kappa|a_t-a^*|+\Phi r_0e^{-gt}.
\end{equation}
Iterating yields $\kappa^t|a_0-a^*|+\Phi r_0\sum_{j=0}^{t-1}\kappa^{t-1-j}e^{-gj}$ as an upper bound. Both terms tend to zero; the same conclusion holds when $\kappa=e^{-g}$, for which the sum is $t\kappa^{t-1}$. Continuity of $\Lam$ and $r_t\to0$ then imply $\ell_t\to\ell^*$.

To see why a positive committed expenditure share can change the destination with three roots, take any $r_0\in(0,1)$ and an initial relative productivity $a_0=a_2^*+\epsilon$, where $\epsilon>0$. With $r_0=0$, that state is in the high basin. Since $F(a_2^*)=a_2^*$ and $\Phi r_0q(a_2^*)>0$, sufficiently small $\epsilon$ gives
\begin{equation}
 F(a_2^*+\epsilon)-\Phi r_0q(a_2^*+\epsilon)<a_2^*.
\end{equation}
The perturbed path then remains strictly below $a_2^*$, because $F$ is increasing and the perturbation is nonpositive. While it is between $a_1^*$ and $a_2^*$, it decreases. If it stays in this interval forever, its limit must be $a_1^*$. If it falls below $a_1^*$, it cannot cross $a_1^*$ from below. On any compact interval bounded away below $a_1^*$, the unperturbed positive drift has a strictly positive minimum, whereas the perturbation tends to zero. The path must therefore eventually enter every neighborhood $[a_1^*-\epsilon,a_1^*]$. Once the uniform perturbation bound $\Phi r_t$ is smaller than $F(a_1^*-\epsilon)-(a_1^*-\epsilon)$, monotonicity of $F$ makes this neighborhood forward invariant. In either case the path converges to the low root. Thus the same initial relative productivity can lead to different stable branches depending on the initial committed expenditure share.
\end{proof}

In the Cobb--Douglas limit, the utility representation becomes, up to a positive normalization conditional on $E$, $(C_L-\gamma)^{1-\alpha}C_H^\alpha$. The resource share for $\gamma=0$ is $\alpha(E)$, so the relative map~\eqref{eq:CD} has slope $1-\delta\in[0,1)$. With $0<\sigma<1$, write $k=\sigma-1<0$. The drift derivative $-\delta+\Phi k\ell(1-\ell)$ is strictly negative and the limiting signs guarantee a unique root. Its local multiplier satisfies $F_a<1$. Local asymptotic stability is guaranteed when
\begin{equation}
 \Phi(1-\sigma)\ell^*(1-\ell^*)<2-\delta.
\end{equation}
The reverse strict inequality implies instability. At equality, the multiplier is $-1$ and linearization alone is inconclusive; nonlinear terms can still produce asymptotic stability. Thus this strict bound is a sufficient condition, not a necessary condition covering the nonhyperbolic boundary.

If catch-up is absent, $\delta=0$, the relative drift is $\nu E-\phi_L+\Phi\Lam(m+ka)$. For $k>0$ this drift is strictly increasing. It has a unique finite zero if and only if $-\phi_H<\nu E<\phi_L$; that zero has multiplier strictly above one. Equality at either bound corresponds to a limiting share of zero or one, not a finite relative productivity. The stable outer branches in the main analysis therefore rely on an explicit convergence force.

\section{Growth and the planner's conditions}\label{app:planner}
\begin{proof}[Proof of Proposition~\ref{prop:growth}]
Stationarity implies $\delta a^*=b_H-b_L$. If $a^*\geq0$, the $H$ catch-up term is zero and the $L$ catch-up term equals $b_H-b_L$. Both growth rates equal $g+b_H$. If $a^*\leq0$, both equal $g+b_L$. The expressions agree at zero. Constant labor shares imply the same log growth for both consumption quantities. Homogeneity of degree one of the CES index at $\gamma=0$ gives the same growth for $U$. Differentiating the two pieces at fixed $E$ gives the stated slope signs.
\end{proof}

The growth ranking can reverse across the two stable configurations. Set $\delta=0.05$, $\phi_H=0.08$, $\phi_L=0.12$, $\sigma=3$, $\nu=0$, $g=0.005$, and $m(E)=0.8$ at the education level considered. Then $K=8$ and $D=-4$. The stable shares are approximately $0.021248$ and $0.978752$. The low-share configuration has common log growth $0.122450$, while the high-share configuration has log growth $0.083300$. Own-sector learning is positive in both sectors, and both configurations are stable. The configuration with a larger share in $H$ grows more slowly in this economy.

\begin{proof}[Proof of Proposition~\ref{prop:wedge}]
The finite-horizon Bellman equation is
\begin{equation}
 V_t(x_L,x_H;E)=\max_{\ell\in[0,1]}
 \left\{u(x_L,x_H,\ell;E)+\beta V_{t+1}(x_L',x_H';E)\right\}.
\end{equation}
At a fixed current productivity state, $\partial x_H'/\partial\ell=\phi_H$ and $\partial x_L'/\partial\ell=-\phi_L$. Differentiating at an interior optimum yields~\eqref{eq:plannerFOC}. In the competitive economy, the current allocation maximizes $U$ and hence $\log U$ over the same current production possibilities. Its static derivative is zero. Evaluating the derivative of the planner's objective at that allocation therefore leaves precisely~\eqref{eq:wedge}. A strictly positive derivative provides a local improvement direction toward higher $\ell$, and a negative derivative the reverse. This local argument allows a nonconcave continuation objective.
\end{proof}

Define the CES contribution shares at the planner's allocation as
\begin{equation}
 \theta_H=\frac{\alpha^{1/\sigma}C_H^\rho}
 {(1-\alpha)^{1/\sigma}C_L^\rho+\alpha^{1/\sigma}C_H^\rho},\qquad
 \theta_L=1-\theta_H.
\end{equation}
Then $u_{x_H}=\theta_H$, $u_{x_L}=\theta_L$, and $u_\ell=\theta_H/\ell-\theta_L/(1-\ell)$. Where $V_t$ and $V_{t+1}$ are differentiable at the relevant current and optimal next states, the envelope equations for $a>0$ are
\begin{align}
 V_{L,t}&=\theta_L+\beta(1-\delta)V_{L,t+1},\\
 V_{H,t}&=\theta_H+\beta\bigl(V_{H,t+1}+\delta V_{L,t+1}\bigr).
\end{align}
For $a<0$ they are
\begin{align}
 V_{L,t}&=\theta_L+\beta\bigl(V_{L,t+1}+\delta V_{H,t+1}\bigr),\\
 V_{H,t}&=\theta_H+\beta(1-\delta)V_{H,t+1}.
\end{align}
Next-period derivatives are evaluated at the optimal next state, and the terminal condition is $V_{j,T+1}=0$. At the catch-up kink $a=0$ and other nondifferentiable states, directional optimality and envelope conditions replace the smooth formulas. The first-order condition and net learning wedge use the continuation-value derivatives at differentiable states. The nonnegative coefficients allow positive marginal values of knowledge in both sectors, while their weighted difference in~\eqref{eq:wedge} can be positive, negative, or zero.

\section{Numerical methods and sensitivity}\label{app:numerics}

\subsection{Root enumeration at the folds}
We locate the configurations using log expenditure odds, $z=\logit\ell$. The fixed-point equation becomes
\begin{equation}
 h(z;E)=z-K\Lam(z)-D(E)=0.
\end{equation}
Every root lies between $D-1$ and $D+K+1$: the residual is negative at the first bound and positive at the second. For $K>4$, transform the turning-point shares in~\eqref{eq:turning} into $z_-=\logit\ell_-$ and $z_+=\logit\ell_+$. The transformed points that lie inside the bracket divide it into monotone segments. Bisection on each segment locates all simple roots. At the analytical folds, we include the tangent root explicitly. Close to a fold, 80-digit arithmetic distinguishes nearby simple roots from a tangency. The case $K=4,D=-2$ has a single root with cubic tangency. Stability follows from~\eqref{eq:multiplier}.

For paths with a committed consumption requirement, we update both log productivities using~\eqref{eq:techH}--\eqref{eq:techL}. At each date, $r_t=\gamma/A_{L,t}$ and $\ell_t=(1-r_t)\Lam(m+ka_t)$. Thus the same labor allocation enters both sectoral learning equations.

\subsection{Local changes in parameters}
Table~\ref{tab:sensitivity} multiplies each reported parameter, one at a time, by $0.9$ or $1.1$ while holding the others fixed. The baseline $g$ is omitted because it cancels from relative dynamics. The education intercept $m_0$ is negative, so multiplying it by $0.9$ moves it toward zero. The table reports the algebraic fold locations even when they fall outside $E\geq0$.

\begin{table}[htbp]
\centering
\caption{Sensitivity of the two education thresholds}\label{tab:sensitivity}
\small\begin{tabular}{lrrrr}
\toprule
Perturbation & $K$ & $E_-$ & $E_+$ & Width \\
\midrule
Baseline & 8.000 & 0.239 & 1.761 & 1.522 \\
$\delta$ $\times 0.9$ & 8.889 & -0.053 & 1.869 & 1.922 \\
$\delta$ $\times 1.1$ & 7.273 & 0.482 & 1.678 & 1.195 \\
$\phi_H$ $\times 0.9$ & 7.520 & 0.530 & 1.813 & 1.284 \\
$\phi_H$ $\times 1.1$ & 8.480 & -0.056 & 1.713 & 1.768 \\
$\phi_L$ $\times 0.9$ & 7.680 & 0.204 & 1.567 & 1.362 \\
$\phi_L$ $\times 1.1$ & 8.320 & 0.271 & 1.957 & 1.686 \\
$\sigma$ $\times 0.9$ & 6.800 & 0.642 & 1.627 & 0.985 \\
$\sigma$ $\times 1.1$ & 9.200 & -0.154 & 1.907 & 2.060 \\
$\eta$ $\times 0.9$ & 8.000 & 0.257 & 1.897 & 1.640 \\
$\eta$ $\times 1.1$ & 8.000 & 0.223 & 1.644 & 1.421 \\
$\nu$ $\times 0.9$ & 8.000 & 0.246 & 1.813 & 1.567 \\
$\nu$ $\times 1.1$ & 8.000 & 0.232 & 1.712 & 1.480 \\
$m_0$ $\times 0.9$ & 8.000 & 0.082 & 1.604 & 1.522 \\
$m_0$ $\times 1.1$ & 8.000 & 0.396 & 1.918 & 1.522 \\
\bottomrule
\end{tabular}
\figurenote{All rows retain $K>4$ and have three simple roots at $E=1$. Negative lower thresholds lie outside the admissible education domain. The width is $E_+-E_-$ in the algebraic extension of the index.}
\end{table}

\begin{thebibliography}{99}

\bibitem[Alan and Ertac(2018)]{Alan2018}
Alan, Sule, and Seda Ertac. 2018.
``Fostering Patience in the Classroom: Results from Randomized Educational Intervention.''
\textit{Journal of Political Economy} 126(5): 1865--1911.
\href{https://doi.org/10.1086/699007}{doi: \nolinkurl{10.1086/699007}}.

\bibitem[Aoki and Yoshikawa(2002)]{Aoki2002}
Aoki, Masanao, and Hiroshi Yoshikawa. 2002.
``Demand Saturation-Creation and Economic Growth.''
\textit{Journal of Economic Behavior \& Organization} 48(2): 127--154.
\href{https://doi.org/10.1016/S0167-2681(01)00229-3}{doi: \nolinkurl{10.1016/S0167-2681(01)00229-3}}.

\bibitem[Arrow(1962)]{Arrow1962}
Arrow, Kenneth J. 1962.
``The Economic Implications of Learning by Doing.''
\textit{Review of Economic Studies} 29(3): 155--173.
\href{https://doi.org/10.2307/2295952}{doi: \nolinkurl{10.2307/2295952}}.

\bibitem[Bernheim and Rangel(2009)]{Bernheim2009}
Bernheim, B. Douglas, and Antonio Rangel. 2009.
``Beyond Revealed Preference: Choice-Theoretic Foundations for Behavioral Welfare Economics.''
\textit{Quarterly Journal of Economics} 124(1): 51--104.
\href{https://doi.org/10.1162/qjec.2009.124.1.51}{doi: \nolinkurl{10.1162/qjec.2009.124.1.51}}.

\bibitem[Boppart(2014)]{Boppart2014}
Boppart, Timo. 2014.
``Structural Change and the Kaldor Facts in a Growth Model with Relative Price Effects and Non-Gorman Preferences.''
\textit{Econometrica} 82(6): 2167--2196.
\href{https://doi.org/10.3982/ECTA11354}{doi: \nolinkurl{10.3982/ECTA11354}}.

\bibitem[Buera and Kaboski(2012)]{Buera2012}
Buera, Francisco J., and Joseph P. Kaboski. 2012.
``The Rise of the Service Economy.''
\textit{American Economic Review} 102(6): 2540--2569.
\href{https://doi.org/10.1257/aer.102.6.2540}{doi: \nolinkurl{10.1257/aer.102.6.2540}}.

\bibitem[Buringh and van Zanden(2009)]{Buringh2009}
Buringh, Eltjo, and Jan Luiten van Zanden. 2009.
``Charting the `Rise of the West': Manuscripts and Printed Books in Europe, A Long-Term Perspective from the Sixth through Eighteenth Centuries.''
\textit{Journal of Economic History} 69(2): 409--445.
\href{https://doi.org/10.1017/S0022050709000837}{doi: \nolinkurl{10.1017/S0022050709000837}}.

\bibitem[Comin et~al.(2021)Comin, Lashkari, and Mestieri]{Comin2021}
Comin, Diego, Danial Lashkari, and Mart\'{i} Mestieri. 2021.
``Structural Change with Long-Run Income and Price Effects.''
\textit{Econometrica} 89(1): 311--374.
\href{https://doi.org/10.3982/ECTA16317}{doi: \nolinkurl{10.3982/ECTA16317}}.

\bibitem[de Vries(1994)]{DeVries1994}
de Vries, Jan. 1994.
``The Industrial Revolution and the Industrious Revolution.''
\textit{Journal of Economic History} 54(2): 249--270.
\href{https://doi.org/10.1017/S0022050700014467}{doi: \nolinkurl{10.1017/S0022050700014467}}.

\bibitem[Fabian and Dold(2025)]{Fabian2025}
Fabian, Mark, and Malte Dold. 2025.
``Agentic Preferences: A Foundation for Nudging When Preferences Are Endogenous.''
\textit{Behavioural Public Policy} 9(1): 249--269.
\href{https://doi.org/10.1017/bpp.2022.17}{doi: \nolinkurl{10.1017/bpp.2022.17}}.

\bibitem[Foellmi and Zweim\"uller(2006)]{Foellmi2006}
Foellmi, Reto, and Josef Zweim\"uller. 2006.
``Income Distribution and Demand-Induced Innovations.''
\textit{Review of Economic Studies} 73(4): 941--960.
\href{https://doi.org/10.1111/j.1467-937X.2006.00403.x}{doi: \nolinkurl{10.1111/j.1467-937X.2006.00403.x}}.

\bibitem[Herrendorf et~al.(2014)Herrendorf, Rogerson, and Valentinyi]{Herrendorf2014}
Herrendorf, Berthold, Richard Rogerson, and \'{A}kos Valentinyi. 2014.
``Growth and Structural Transformation.''
In \textit{Handbook of Economic Growth}, vol.~2, edited by Philippe Aghion and Steven N. Durlauf, 855--941. Elsevier.
\href{https://doi.org/10.1016/B978-0-444-53540-5.00006-9}{doi: \nolinkurl{10.1016/B978-0-444-53540-5.00006-9}}.

\bibitem[Kongsamut et~al.(2001)Kongsamut, Rebelo, and Xie]{Kongsamut2001}
Kongsamut, Piyabha, Sergio Rebelo, and Danyang Xie. 2001.
``Beyond Balanced Growth.''
\textit{Review of Economic Studies} 68(4): 869--882.
\href{https://doi.org/10.1111/1467-937X.00193}{doi: \nolinkurl{10.1111/1467-937X.00193}}.

\bibitem[Lucas(1988)]{Lucas1988}
Lucas, Robert E., Jr. 1988.
``On the Mechanics of Economic Development.''
\textit{Journal of Monetary Economics} 22(1): 3--42.
\href{https://doi.org/10.1016/0304-3932(88)90168-7}{doi: \nolinkurl{10.1016/0304-3932(88)90168-7}}.

\bibitem[Matsuyama(2002)]{Matsuyama2002}
Matsuyama, Kiminori. 2002.
``The Rise of Mass Consumption Societies.''
\textit{Journal of Political Economy} 110(5): 1035--1070.
\href{https://doi.org/10.1086/341873}{doi: \nolinkurl{10.1086/341873}}.

\bibitem[Michael(1973)]{Michael1973}
Michael, Robert T. 1973.
``Education in Nonmarket Production.''
\textit{Journal of Political Economy} 81(2, Part~1): 306--327.
\href{https://doi.org/10.1086/260029}{doi: \nolinkurl{10.1086/260029}}.

\bibitem[Stigler and Becker(1977)]{Stigler1977}
Stigler, George J., and Gary S. Becker. 1977.
``De Gustibus Non Est Disputandum.''
\textit{American Economic Review} 67(2): 76--90.
\href{https://www.jstor.org/stable/1807222}{JSTOR: \nolinkurl{1807222}}.

\bibitem[Woodford(2022)]{Woodford2022}
Woodford, Michael. 2022.
``Effective Demand Failures and the Limits of Monetary Stabilization Policy.''
\textit{American Economic Review} 112(5): 1475--1521.
\href{https://doi.org/10.1257/aer.20201529}{doi: \nolinkurl{10.1257/aer.20201529}}.

\end{thebibliography}
\end{document}